# NLP-assisted software testing: A systematic mapping of the literature

Vahid Garousi
Queen's University Belfast
Northern Ireland, UK
v.garousi@qub.ac.uk

Sara Bauer
University of Innsbruck,
Austria
sara.bauer@uibk.ac.at

Michael Felderer
University of Innsbruck, Austria
Blekinge Institute of Technology, Sweden
michael.felderer@uibk.ac.at

**Abstract**:

*Context*: To reduce manual effort of extracting test cases from natural-language requirements, many approaches based on Natural Language Processing (NLP) have been proposed in the literature. Given the large amount of approaches in this area, and since many practitioners are eager to utilize such techniques, it is important to synthesize and provide an overview of the state-of-the-art in this area.

*Objective*: Our objective is to summarize the state-of-the-art in NLP-assisted software testing which could benefit practitioners to potentially utilize those NLP-based techniques. Moreover, this can benefit researchers in providing an overview of the research landscape.

*Method*: To address the above need, we conducted a survey in the form of a systematic literature mapping (classification). After compiling an initial pool of 95 papers, we conducted a systematic voting, and our final pool included 67 technical papers.

*Results*: This review paper provides an overview of the contribution types presented in the papers, types of NLP approaches used to assist software testing, types of required input requirements, and a review of tool support in this area. Some key results we have detected are: (1) only four of the 38 tools (11%) presented in the papers are available for download; (2) a larger ratio of the papers (30 of 67) provided a shallow exposure to the NLP aspects (almost no details).

*Conclusion*: This paper would benefit both practitioners and researchers by serving as an "index" to the body of knowledge in this area. The results could help practitioners utilizing the existing NLP-based techniques; this in turn reduces the cost of test-case design and decreases the amount of human resources spent on test activities. After sharing this review with some of our industrial collaborators, initial insights show that this review can indeed be useful and beneficial to practitioners.

# Table of Contents

# 1 Introduction

Software testing is a fundamental activity to ensure a certain degree of quality in software systems. However, testing is an effort-intensive activity. In its conventional form, human testers (test engineers) conduct most (if not all) phases of software testing manually. One of those phases is test-case design in which the human tester uses written (formal) requirements, written often in natural language (NL), to derive a set of test cases. Test-case design is also an effort-intensive activity [1], and practitioners are eager to get help from any (partially) automated approach to extract test suites from requirements [1]. Such a practice could save software companies considerable resources which are regularly spent to manually derive and document test cases from requirements. Furthermore, as software requirements change, test cases have to be maintained, an activity which incurs further effort.

To reduce the manual effort of converting natural-language (NL) requirements into test cases, many approaches based on Natural Language Processing (NLP) have been proposed in the literature. Such an approach requires an input set of requirements written in NL. Then, following a series of NLP steps [2], a set of test cases are extracted automatically from the textual requirements. Let us clarify that we use the conventional definition of a "test case" [3] in this work: a test case is one or more inputs (as needed) and the excepted output(s) (or behavior) for a unit or a system under test. For example, to test an absolute-value function, one would need at least two test cases: a test case with a positive integer, and another test case with a negative integer.

In addition to the test-case design phase, NLP techniques have also been used in other software testing activities, e.g., in the context of the test oracle problem, e.g., [4].

To improve the efficiency of software testing, many NLP-based techniques and tools have been proposed in the last decades. We use the phrase "NLP-assisted software testing" in this paper to refer to all NLP-based techniques and tools which could assist any software testing activity, e.g., test-case design and test evaluation, as discussed above.

Given the growing body of knowledge in the area of NLP-assisted software testing, reviewing and getting an overview of the entire state-of-the-art and -practice in this area is challenging for a practitioner or a (new) researcher. As discussed above, practitioners are eager to get help from any (partially) automated approach to help them save time in extracting tests from requirements [1]. Knowing that they can adapt/customize an existing technique to predict and improve software testing in their own context can potentially help companies and test engineers bring more efficiency into their software testing practices. Thus, we have observed first-hand that there is a real need for review papers like the current one to provide a summary of the entire field and serve as an "index" to the body of knowledge in this area, so that a practitioner can get a snapshot of the current knowledge without having to find and read through all of the papers in this area. Furthermore, a recent insightful paper in IEEE Software [5] highlighted "*the practical value of historical data [and approaches published in the past]*" and a "*vicious cycle of inflation of software engineering terms and knowledge*" (due to many papers not adequately reviewing the state of art). We believe survey papers like the current one aim at addressing the above problem.

To systematically review and get an overview of studies in a given research area, Systematic Literature Review (SLR) and Systematic (Literature) Mapping (SLM or SM) studies are the established approaches. To address the above need and to find out what we, as a community, know about NLP-assisted software testing, we report in this paper a SLM in this area. Our review pool included 67 academic peer-reviewed papers. The first paper in this area was published in 2001 and this review study includes all the papers until end of 2017. A few review (survey) papers have been previously published in this area, e.g., [6, 7], but their review pools were somewhat limited as the largest paper pool size amounted to 16 papers (in [7]). As we discuss in Section 2.3, our survey is the most up-to-date and comprehensive review in the area by considering all the 67 papers, that we have found, published in this area between 2001-2017.

The remainder of this paper is structured as follows. Background and related work is presented in Section 2. We describe the research method and the planning phase of our review in Section 3. Section 4 presents the results of the literature review. Section 5 summarizes the findings and potential benefits of this review. Finally, in Section 6, we draw conclusions, and suggest areas for further research. In the appendix, we show the list of the primary studies reviewed in this survey.

## 2 BACKGROUND AND RELATED WORK

In this section, we first provide a brief overview of the concept of NLP, followed by an overview of NLP-assisted software testing. We then review the related work, which are the existing survey (review) papers on NLP-assisted software testing.

### 2.1 AN OVERVIEW OF NLP

NLP is a prominent sub-field of both computational linguistics and artificial intelligence (AI). NLP covers the "*range of computational techniques for analyzing and representing naturally-occurring texts [...] for the purpose of achieving human-like language processing*" [8]. Challenges in NLP usually involve speech recognition, natural-language understanding, and natural-language generation.

NL structures might be rule-based from a syntactic point of view, yet the complexity of semantics is what makes language understanding a rather challenging idea. For instance, a study reported that the sentence "*List the sales of the products produced in 1973 with the products produced in 1972.*" offered 455 different semantic-syntactic parses [8]. This clearly demonstrates the problems of computational processing: while linguistic disambiguation is an intuitive skill in humans, it is difficult to convey all the small nuances that make up NL to a computer. For that reason, different sub-fields of NLP have emerged to analyze aspects of NLP from different angles. Those sub-fields include: (1) Discourse Analysis [9], a rubric assigned to analyze the discourse structure of text or other forms of communication; (2) Machine Translation [10], intended to translate a text from one human language into another, with popular tools such as "Google Translate"; and (3) information extraction (IE), which is concerned with extracting information from unstructured text utilizing NLP resources such as lexicons and grammars [11]. Among all NLP approaches, IE is often the most widely used in the software engineering context [12]. We thus provide an overview of concepts in IE.

IE is described by three dimensions: (1) the structure of the content plays a role, ranging from free text, HTML, XML, and semi-structured NL; (2) the techniques used for processing the text must be determined; and (3) the degree of automation in the collecting, labeling and extraction process must be considered. For structured text such as HTML or XML, information retrieval is delimited by the labels or tags, which can be extracted. Free text, however, requires a much thorough analysis prior to any extraction. In the following, we briefly explain the concepts from IE, which are relevant for this paper.

- Morphology:
  - Part of Speech Tagger (POS): A form of grammatical tagging in which a phrase (sentence) is classified according to its lexical category. The categories include nouns, verbs, adverbs, adjectives, pronouns, conjunction and their subcategories.
  - Stemming: In stemming, derived words are reduced to their base or root forms. For example, the words "am, is, are" are converted to their root form "be".
  - Named-Entity Recognition (NER): NER allocates types of semantics such as *person*, *organization* or *localization* in a given text [13].
- Syntax:
  - Constituency/Dependency Parsing: Although sometimes used interchangeably, Dependency Parsing focuses on the relationships between words in a sentence (in its simplest form, the classic subject-verb-object structure). Constituency Parsing, however, breaks a text into sub-phrases. Non-terminals in the parse tree are types of phrases (noun or verb phrases), whereas the terminals are the words in the sentence, yielding a more nested parse tree.
- Semantics:
  - Semantic Role Labeling (SRL): SRL is also called shallow semantic parsing. In SRL, labels are assigned to words or phrases in a sentence that indicate their semantic role in the sentence. These roles can be agent, goal, or result.
  - Word Sense Disambiguation: Detecting the correct meaning of an ambiguous word used in a sentence.

The above definitions shall help the reader understand the NLP concepts, and their usage in software testing, when reading the rest of this paper. As a concrete example, we show in Figure 1 the outputs of applying several NLP techniques on the following example NL requirement item: `If the user enters valid user name and password, then the system should let the user log in`. We have used two online tools to do this example analysis: www.corenlp.run and macniece.seas.upenn.edu:4004.

| | |
|---|---|
| POS: | [IN] [DT] [NN] [VBZ] [JJ] [NN] [NN] [CC] [NN] [,] [RB] [DT] [NN] [MD] [VB] [DT] [NN] [NN] [IN] [.]<br>If the user enters valid user name and password , then the system should let the user log in . |
| SRL: | A0 — Predicate — A1 — A0 — AM-MOD — Predicate<br>If the user enters valid user name and password , then the system should let the user log in . |
| Information extraction (IE): | Entity — subject — Relation — object — Entity — object — Entity — Entity — subject — Relation — object — Entity<br>If the user enters valid user name and password , then the system should let the user log in . |
| Basic dependency parsing: | [IN] [DT] [NN] [VBZ] [JJ] [NN] [NN] [CC] [NN] [,] [RB] [DT] [NN] [MD] [VB] [DT] [NN] [NN] [IN] [.]<br>If the user enters valid user name and password , then the system should let the user log in . |

**Figure 1- Applying several NLP techniques on an example NL requirement item**

## 2.2 A BRIEF OVERVIEW OF NLP-ASSISTED SOFTWARE TESTING

Software testing is an important but effort-intensive activity. In its conventional form, human testers (test engineers) conduct most or all activities of software testing manually. One of those activities is test-case design in which the human tester uses formal or informal requirements to derive and (sometimes) document test suites (set of test cases).

To reduce the cost of various activities of software testing, test automation has become popular in the last decades. This helps reducing the manual work-load in various software testing activities, see for example [14, 15]. Test cases are developed to drive the execution of software items and comprise preconditions, inputs (including actions, where applicable), and expected outputs.

To reduce the manual effort in converting requirements, written in NL, to test cases, many approaches based on Natural Language Processing (NLP) have been proposed in the literature. Such approaches provide test automation for the test-case design phase. Such an approach requires an input set of requirements to be written in NL. Then following a series of NLP [2] steps (also called NLP "pipeline"), a set of test cases is extracted automatically from the textual requirements. Of course, the precision of such a transformation is usually not perfect and often needs peer review by a human tester [16, 17] (also see Section 4.3.4). To help the reader who is not familiar with NLP and different steps for NLP [2], we depict the classical NLP pipeline of steps in Figure 2 (taken from [2]).

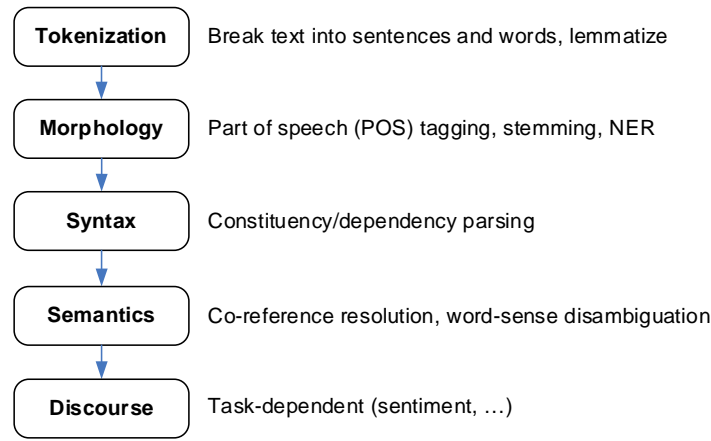

**Figure 2- Classical NLP Pipeline [2]**

To show the concept of NLP-assisted software testing, we show in Table 1 two example NL requirement items for the "Login" use-case of a typical web application. An approach to enable NLP-assisted test-case design, in this context, would take these requirement items as input, and perhaps also use other information such as the system's context and class diagrams, to generate a set of test cases as shown in Table 1. As we can see in Table 1, the two requirement items mention that user name and password combinations could be valid or invalid, but concrete values of such cases are often not documented directly in the requirements document and such data could be automatically extracted from other sources (e.g., account databases or UML diagrams of the system). A typical NLP-assisted test-case design approach would generate as output the list of test cases based on the NL requirements.

**Table 1: An example showing the concept of NLP-assisted software testing**

| Inputs: Requirement items written in natural language | Outputs: Test cases | |
|---|---|---|
| | Inputs | Expected output / system state |
| • If the user enters valid user name and password (such as "user", "password"), then the system should let the user log in. | User name="user", Password="password" | user_session =logged_in |
| • If the user does not enter valid user name and password, then the system should not let the user log in, and should show this error message: "Incorrect username / password". | User name="user", Password="incorrect" | user_session=not_logged_in AND page.contains("Incorrect username / password") |

An important issue for NLP-assisted software testing is the type of NL requirements taken as input by a given approach. As we will review in this survey paper (Section 4.2.3), while some approaches need the software requirements be expressed in restricted (controlled) NL [2], some other approaches allow for more freedom in the way the input requirements are written. Controlled NL is a subset of NL that is obtained by restricting the grammar and vocabulary [2], in order to reduce or eliminate ambiguity and complexity of the NLP-based technique for extracting test cases from the requirements.

## 2.3 RELATED WORK: OTHER SURVEY (REVIEW) PAPERS IN THIS AREA

A few survey/review papers (secondary studies) have been reported in this area. We were able to find four such studies [6, 7, 18, 19] and provide their list in Table 2. For each study, we include its publication year, its type (regular survey or systematic mapping/review), number of papers reviewed by the study, and some explanatory notes. As we can see in Table 2, three of those four review papers were not "focused" on NLP-based test generation papers, but instead, those papers were only a subset of their review pools. Only one of these papers [7] was focused on NLP-based test generation which compiled a set of six NLP-based techniques and 18 tools. However, our review provides a list of 67 techniques and 38 associated tools. Therefore, our survey is the most up-to-date and comprehensive review in the area by considering all 67 papers in this area, published between 2001-2017.

Another remotely-related work is a 2017 SLR on applications of NLP in software requirement engineering [20], which reviewed a pool of 27 papers. However, the paper did not focus on NLP-assisted software testing.

**Table 2: A list of survey (review) papers on NLP-assisted software testing**

| Paper title | Year | Reference | Type of study | Num. of papers reviewed | Notes |
|---|---|---|---|---|---|
| Test case derived from requirement specification | 2003 | [18] | Regular survey | 55 papers (only 9 papers derived test cases from requirements in natural language) | Not focused on NLP. NLP-based test generation papers were a subset of the pool |
| Generation of test cases from functional requirements: a survey | 2006 | [19] | Regular survey | 13 approaches (papers) | Not focused on NLP |
| An overview on test generation from functional requirements | 2011 | [6] | Regular survey | 22 papers in total (only 1 NLP-based approach) | Not focused on NLP |
| A comprehensive investigation of natural language processing techniques and tools to generate automated test cases | 2017 | [7] | SLR | 16 papers | 6 NLP-based techniques and 18 tools |
| This study | 2018 |  | SLM | 67 papers | We compile a list of 38 tools |

## 3 RESEARCH METHOD (PLANNING OF THE SYSTEMATIC REVIEW)

Based on our past experience in SLR and SLM studies, e.g., [21], and also using the established guidelines for conducting SLR and SLM studies in SE (e.g., [22-25]), we developed our review process, as shown in Figure 3. We discuss the planning and design phases of our review in the next sections.

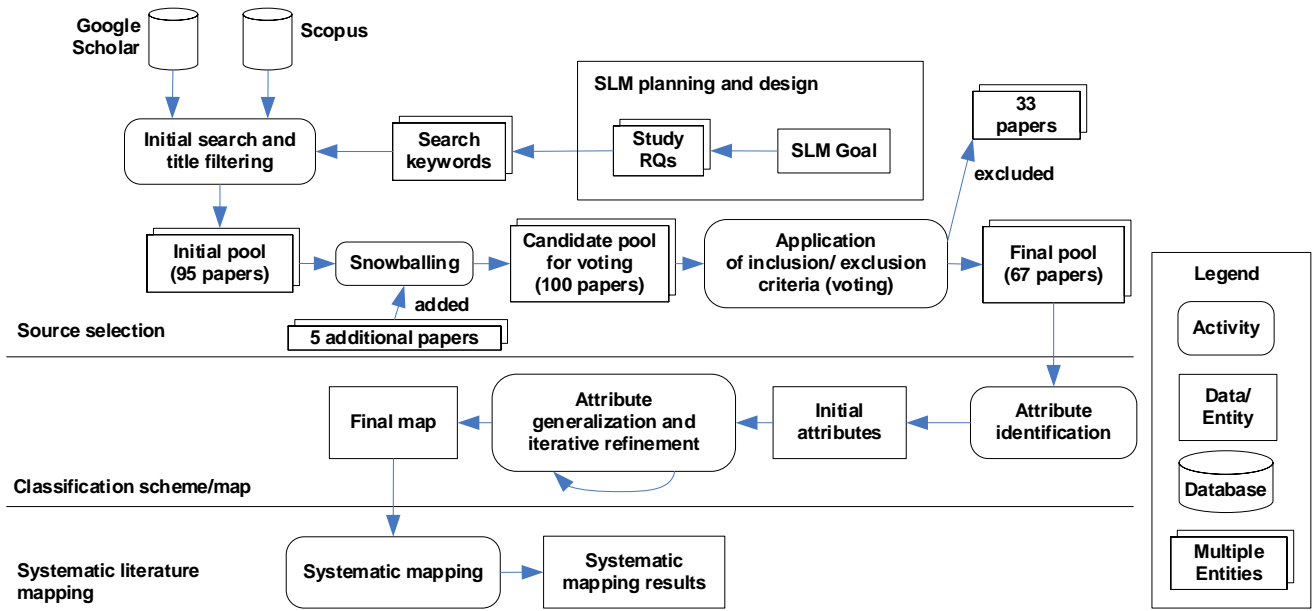

**Figure 3-An overview of our review process (as a UML activity diagram)**

## 3.1 GOAL AND REVIEW QUESTIONS

The goal of this study is to systematically map (classify), review and synthesize the state-of-the-art in the area of NLP-assisted software testing. Moreover, this study aims at detecting recent trends and directions in this field, and aims at identifying opportunities for future research, from both researchers' and practitioners' perspectives. Based on the above goal, we raise the following review questions (RQs), which we group under three categories:

Group 1-Common to all SLM studies:

- **RQ 1.1-Mapping of studies by contribution types:** What are the different types of contributions presented in the papers? How many papers have presented approaches, methods, or tools for NLP-assisted software testing?
- **RQ 1.2-Mapping of studies by research-method types:** What type of research methods have been used in the papers? Some of the papers presented solution proposals while other papers used more rigorous research methods such as empirical studies.

Group 2-Specific to the topic (NLP-assisted software testing):
- **RQ 2.1-Type of NLP approaches used to assist software testing:** What type of NLP approaches have been used to assist software testing? Examples of popular NLP approaches include: morphology, syntactic and semantic analysis [2].
- **RQ 2.2-Exposure level of the NLP aspects (algorithm) in the paper:** To what extent has each paper presented the details of the NLP aspects (algorithm)? This RQ is important as some papers presented (almost) all details of the presented NLP algorithm (in-depth), while some other papers have exposed those aspects in a (very) shallow manner (almost no details). Our motivation for this RQ is that if a researcher or a practitioner wants to adopt and implement an NLP algorithm, presented in a paper, s/he would need (almost) all algorithmic details (e.g., the NLP "pipeline" [2]) to develop it, or s/he cannot easily implement/use it.
- **RQ 2.3-Type of input NL requirements:** What type of NL requirements does each approach require as input? While some approaches allow for requirements in "unrestricted" NL, some others can only process requirements in "restricted" NL (i.e., using only a predefined set of keywords in the requirements).
- **RQ 2.4-Intermediate model type (if any):** What types of intermediate model have been created in the NLP-assisted software testing approach? In our initial screening, we have seen some papers that do not transform NL requirements to test cases directly, but instead create an "intermediate" model (such as UML state machines) and then derive test cases from those models.
- **RQ 2.5-Tool support (tool presented):** What tools have been presented in the papers? And what ratio of those tools are publicly available for download? There has been a recent discussion trend in the research community in general about the importance of making research tools available which could lead to various benefits, e.g., reproducible research [26-28].
- **RQ 2.6-NLP tool(s) used:** What NLP tools have been used in the papers? We were curious to see which NLP tools are popular in this area, e.g., the Stanford Parser [29], Natural Language Toolkit (NLTK) [30].
- **RQ 2.7-Language support (under observation):** What (natural) languages are supported for requirements in the papers? We have seen that, while most papers considered requirements in English, a few papers presented approaches for other languages, such as Japanese and German.
- **RQ 2.8-Output of the technique**: What types of test artifacts are generated in each paper? While some papers use NLP to generate test cases (test inputs), some other papers generate other test artifacts such as test oracles.

Group 3-Specific to primary studies which have included empirical case studies:

- **RQ 3.1-Research questions studied by each of the empirical studies**: What are the research questions raised and studied in the empirical studies? Answering this RQ will assist us and readers (e.g., younger researchers) in detecting the types of empirical issues explored in this area so far, and to come up with potential interesting future research directions.
- **RQ 3.2- Scale (size metrics) of the case study:**
    - How many Systems Under Test (SUTs) (or hypothetical examples) have been evaluated in each paper? One would expect that each paper applies the proposed technique to at least one SUT. Some papers take a more comprehensive approach and apply the proposed testing technique to more SUTs.
    - How many requirements items have been processed by each NLP approach?
    - How many test cases have been generated by each NLP approach?
- **RQ 3.3-Methods used to evaluate the NLP approach and empirical evidence:** Which types of methods have been used to evaluate the proposed NLP approach and what is the reported empirical evidence?

- **RQ 3.4-Accuracy (precision) of the approaches**: What are the reported accuracy scores of the presented NLP-based test-case generation approach? This is a follow-up to RQ 3.3. For readers who could potentially consider applying an NLP-based approach, accuracy (precision) of the approach is important since they would want to know how effective the approach is; for instance, what ratio of manual test cases could be generated by the automated approach?

## 3.2 SEARCHING FOR AND SELECTION OF PAPERS

Let us recall from our review process (Figure 3) that the first phase of our study is the selection of papers. For this phase, we followed the following steps, as discussed next:

- Source selection and search keywords
- Application of inclusion and exclusion criteria

### 3.2.1 Selecting the source engines and search process

We selected the source engines and conducted the search process using the established process for performing SLR studies in software engineering, and the established guidelines [22-25]. We performed the searches in both the Google Scholar database and Scopus (www.scopus.com), both of which are widely used in review studies [22-25]. The reason that we used Scopus in addition to Google Scholar was that several papers have mentioned that: "*it [Google Scholar] should not be used alone for systematic review searches*" [31] as it may miss to find some papers.

Using guidelines for screening of primary studies for systematic reviews, e.g., [32, 33], and using our past experience in SLM/SLR studies, e.g., [34], we developed our search strategy and search strings in an experimental manner. The general suggestion (from the above studies) is to design the search terms such that they are not too narrow, and not too broad. This is to ensure maximizing chances of including all relevant studies, while not having to deal with many unrelated papers (false positives) during the search/screening process.

Our search string for Google Scholar was: "*Software AND (test OR testing) AND (natural language OR NLP OR natural language processing)*". When we used other similar terms, e.g., "processing of plain language" instead of NLP in the above search string, our search results returned many unrelated papers (false positives) which would make the filtering task tedious.

We executed the above search string in Scopus and retrieved an additional 15 candidate papers. Furthermore, to ensure maximizing chances to include all relevant papers in Scopus, we experimentally designed, in case of Scopus, an additional search string as: *( TITLE-ABS-KEY ( ( test OR testing ) AND ( "natural language" OR nlp OR "natural language processing" ) ) AND SRCTITLE (software) )*. The *SRCTITLE(software)* term limited the search scope to only venues (journal or conference names) which include the term "software", and which was found to be an effective way to search for software engineering papers in one of our previous studies [35]. Searching by the latter search string retrieved a set of 26 additional candidate papers.

All the authors did independent searches using the search string. In terms of timeline, the search phase was conducted during March 2019, but we only included papers published until end of 2017, since it takes a while for all papers to be actually included in the databases. Data extraction from the primary studies and their classifications were conducted during the same period.

To balance precision, rigor and efficiency in our paper search and selection process, we already conducted title and abstract filtering to ensure that we would add to our candidate paper pool only those papers which are directly or potentially relevant. We had followed the same heuristic in our past SLR and SM studies, e.g., [34]. After all, it would have been meaningless to add a clearly irrelevant paper to the candidate pool and then remove it by application of inclusion/exclusion criteria (Section 4.2). Our first inclusion/exclusion criterion (discussed in Section 4.2) was used for this purpose (i.e., Does the paper focus on NLP-assisted software testing?). For example, Figure 4 shows a screenshot of our search activity using Google Scholar in which potentially relevant candidate papers are highlighted in green, while clearly irrelevant candidate papers are highlighted in red. To ensure efficiency of our efforts, we only added potentially relevant candidate papers to the initial pool.

Another issue was the stopping condition when searching using Google Scholar. We observed that Google Scholar provided a large number of hits (more than 2 million records) using the above keyword, when the search phase was executed. Going through all of them was simply impossible for us. To cope with this challenge, we utilized the relevance ranking of the search engine (Google's PageRank algorithm [36]) to restrict the search space. The good news was that, as per our observations, relevant results usually appeared in the first few pages and as we went through the pages, relevancy of results decreased. Thus, we checked the first n pages (i.e., somewhat a search "saturation" effect) and only continued further if needed, e.g., when at least one result in the n[th] page still was relevant (if at least one paper focused on NLP-assisted testing). Similar heuristics have been reported in several other review studies, be it guideline or experience papers [37-40]. At the end of our initial search and title filtering, our initial pool consisted of 95 papers (as shown in Figure 3).

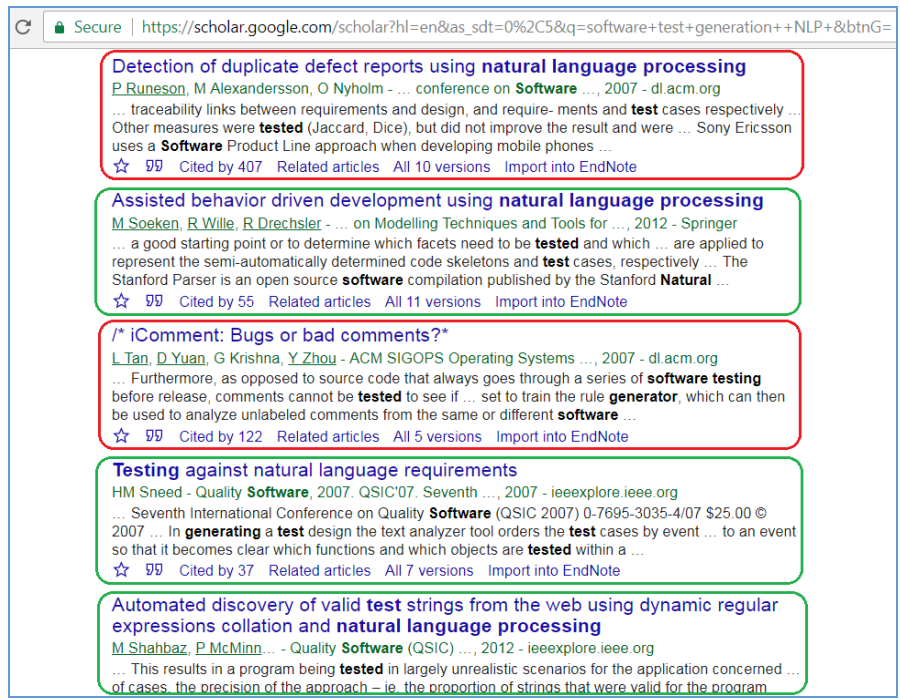

Figure 4- A screenshot from the search process using Google Scholar. Potentially relevant candidate papers are highlighted by green, while clearly irrelevant candidate papers are highlighted by red.

To maximize our search coverage (reach) of all relevant papers as much as possible, we also conducted forward and backward *snowballing* [23] on the papers already in the pool, as is recommended by systematic review guidelines. Snowballing, in this context, refers to using the reference list of a paper (backward snowballing) or the citations to the paper to identify additional papers (forward) [23].

Via snowballing, we found five (5) additional papers, increasing the pool size of the candidate papers to 100. Two examples of the papers found via snowballing are the following. We found [P5] by "forward" snowballing from [P10]. We also found [P32] by forward snowballing of [P46]. Note that, throughout the rest of this paper, we will refer to each of the primary studies by using this format: [Pi], where *i* is the sequential ID of the paper in the pool, e.g., [P4]. They are available in the online dataset of this study: goo.gl/VE6FeK [41], and are also listed in the appendix.

After compiling an initial pool of 100 "candidate" papers, a systematic voting (as discussed next) was conducted among the authors, in which a set of defined inclusion/exclusion criteria were applied to derive the final pool of the primary studies.

### 3.2.2 Application of inclusion/exclusion criteria and voting

We carefully defined the inclusion and exclusion criteria to ensure including all the relevant papers and not including the out-of-scope papers. The inclusion criteria were as follows:

1. Does the paper focus on NLP-assisted software testing?
2. Does the paper include a relatively sound validation?
3. Is the source in English and can its full-text be accessed on the internet?

The answer for each question could be either Yes (value=1) or No (value=0). We included only those papers which received 1's for all criteria, and excluded the rest. Application of the above criteria led to exclusion of 33 papers, details for which can also be found in the study's online dataset [41]. For example, we excluded [42] since only its title was in English, while the paper body was in Portuguese.

## 3.3 FINAL POOL OF THE PRIMARY STUDIES

We finalized the pool with 67 papers. To analyze the growth of this area, we depict in Figure 5 the annual number of papers by their publication years. Note that, as discussed in Section 4.1, we included papers published until the end of 2017. As visualized in Figure 5, the annual number of papers in this area reached its peak in 2017 (10 papers).

Comparing annual publication trends of different topics in software engineering has been reported in many studies, e.g., in a recent 2018 paper in IEEE Transactions on Software Engineering [43]. Such an analysis could provide insights about levels of attention in the research community on different topics. Since we had access to available data from several previous SLM/SLR studies in different areas of software testing, we could easily do such a trend comparison, as shown in Figure 5. The other five topics are: (1) a SLM on web application testing [44], (2) a SLM on testing embedded software [21], (3) a SLM on Graphical User Interface (GUI) testing [45], (4) a SLM on software testability [46], and (5) a survey on mutation testing [47].

Note that the data for the other areas do not reach up to 2017, since the execution and publication timelines of those survey papers are in earlier years. For instance, the survey on mutation testing [47] was published in 2011 and thus only encompasses data up until 2009. Still, the figure provides a reasonable comparative view of the growth of these six sub-areas of software testing.

As we can see in Figure 5, the NLP-assisted software testing area has not been too active when compared to the other areas. However, this area is getting more active in recent years, especially when compared to software testability. The earliest paper in this area was published in 2001 and until year 2010, the papers on this topic were published in a "sporadic" fashion. Such an observation could have a variety of justifications, e.g., perhaps researchers in testing were not that keen to use NLP approaches until quite recently. However, the fact that there are now 67 papers in this topic warrants attention to this topic and to provide a synthesized summary of the topic.

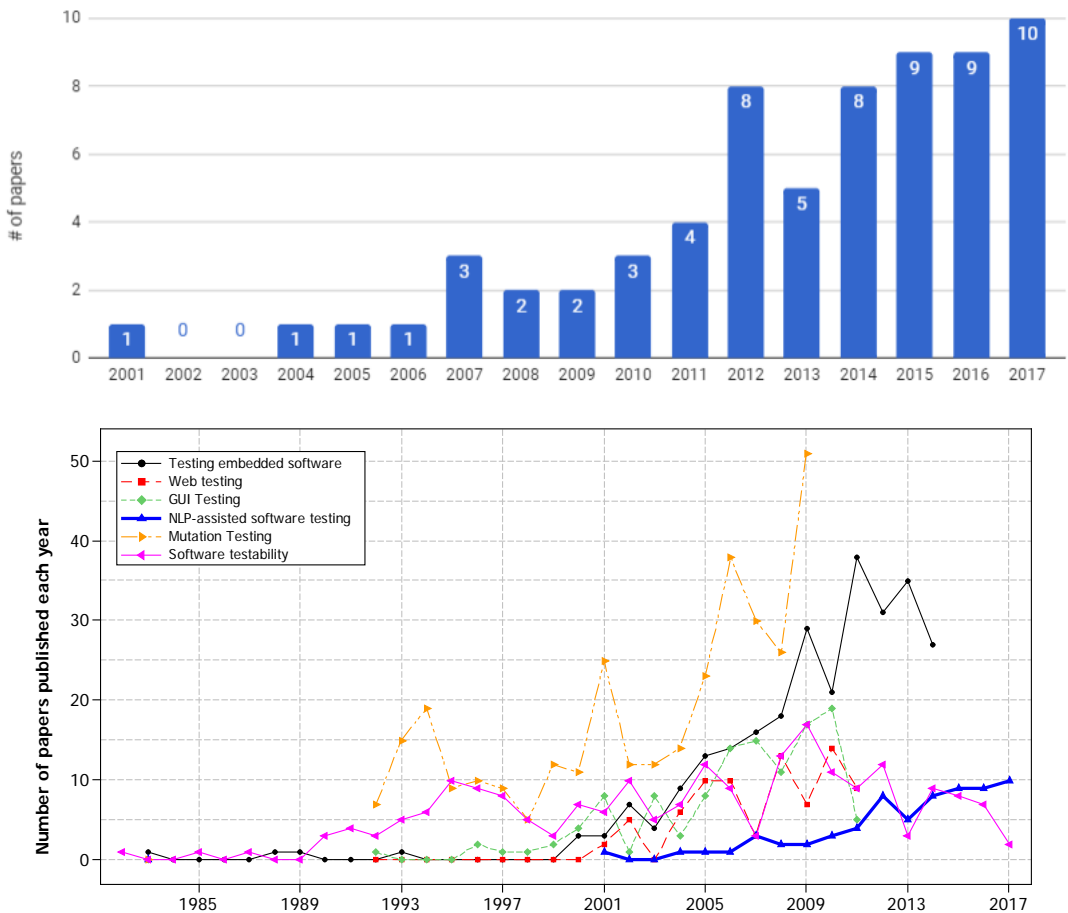

**Figure 5-Growth of the "NLP-assisted software testing" area (top) and comparing the trend of publications in this area with five other software testing topics (bottom)**

## 3.4 Development of the systematic map and data-extraction plan

To answer each of the RQs, we developed a systematic map and then extracted data from papers to classify them using it. We discuss next how we developed the systematic map.

To develop the systematic map, we analyzed the studies in the pool and identified the initial list of attributes. As shown in Figure 3, we then used attribute generalization and iterative refinement, when necessary, to derive the final map.

As papers were identified as relevant to our study, we recorded them in a shared spreadsheet to facilitate further analysis. Our next goal was to categorize the studies in order to gain a holistic impression of the research area and to answer the study RQs. We refined these broad interests into the systematic map using an iterative approach.

Table 3 shows the final classification scheme that we developed after applying the process described above. In the table, column 2 is the list of RQs, column 3 is the corresponding attribute/aspect. Column 4 describes categories / metrics. Column 5 indicates for each attribute whether multiple selections can be applied. For example, in RQ 1.2 (research type), the corresponding value in the last column is 'S' (Single). This indicates that we can classify a given source (paper) under only one research type. In contrast, for RQ 1.1 (contribution type), the corresponding value in the last column is 'M' (Multiple). It indicates that a study can contribute to more than one option (e.g. method, tool, etc.). Classifications of contribution type and research type in Table 3 were done using the well-known guidelines for conducting SLR and SLM studies [22-25].

We derived the categories/metrics for contribution and research types from the SLM guideline paper by Petersen et al. [22]. The contribution types could be: Approach (method, technique), tool, model, metric, process, empirical results only, "other" types. A paper could present (contribute) more than one of the above types, e.g., a paper can present a new technique, and a (prototype) tool to support automating the technique. The contribution of some studies are empirical results only, e.g., [P4], which reported an industrial study to analyze the performance of three existing NLP-based test-case prioritization techniques in the context of 30 industrial projects.

**Table 3: Systematic map developed and used in our study**

| Group | RQ | Attribute/Aspect | Categories/metrics | (M)ultiple/ (S)ingle |
|---|---|---|---|---|
| Group 1-Common to all SLM studies | 1.1 | Contribution type | Approach (method, technique), tool, model, metric, process, empirical results only, other | M |
| | 1.2 | Research type | Solution proposal (proof of concept), weak empirical study (validation research), strong empirical study (evaluation research), experience studies, other | S |
| Group 2-Specific to the topic (NLP-assisted software testing) | 2.1 | Type of NLP approaches to assist software testing | Morphology (POS, stemming, NER, other), Syntax (other, semantic role labeling, co-reference resolution, word-sense disambiguation, other), Other NLP technique used | M |
| | 2.2 | Exposure level of NLP aspects (algorithm) | Very shallow (almost no details), average (few details), in-depth (most details) | S |
| | 2.3 | Type of input NL requirements | Unrestricted NL, restricted NL, Other | S |
| | 2.4 | Intermediate model type | Any intermediate model type | S |
| | 2.5 | Tool support | Available or not available | S |
| | 2.6 | NLP tool used | Stanford Parser, NLTK, Other | M |
| | 2.7 | Language under observation | English, Japanese, Other | M |
| | 2.8 | Output of the technique | Test cases, test oracles, artifacts in support of testing, other | M |
| Group 3-Specific to empirical studies and those with case studies | 3.1 | Research questions raised and studied in the empirical studies | The list of RQs | M |
| | 3.2 | Methods used to evaluate the approaches | Categories of evaluation methods as done by qualitative coding [48] in Section 6.3.3 | M |
| | 3.3 | Scale (size metrics) of the case study system | Any size metrics about the case study system, such as: number of systems/ projects discussed in the paper, number of requirements items, or number of test cases produced by the proposed technique | M |
| | 3.4 | Accuracy (precision) of the approaches | The quantitative reported accuracy score(s). If there are multiple values in a single paper, calculate their average | M |

Among the research-method types (which we derived from [22]), the least rigorous type is "Solution proposal" in which a given study only presents a simple example (or proof of concept). We grouped empirical evaluations under two categories: weak empirical studies (validation research) and strong empirical studies (evaluation research). The former holds when the study does not pose any hypotheses or research questions and does not conduct statistical tests (e.g., using t-test). We considered an empirical evaluation "strong" when it has considered these aspects. Explanations (definitions) of experience studies, philosophical studies, and opinion studies are provided in Peterson et al.'s guideline paper [24].

As for the types of NLP approaches used (RQ 2.2), we followed the classification which is often implemented in NLP pipelines (see Figure 1): Morphology (POS, stemming, NER, other), Syntax (other, semantic role labeling, co-reference resolution, word-sense disambiguation, other), and added Other NLP technique used.

The exposure level of the NLP algorithm, used or presented in a given paper, was classified as: (1) very shallow: if almost no details about the NLP algorithm (approach) were presented; (2) average exposure level: if the description included some details about the underlying approach; and (3) in-depth exposure level: if sufficient details about the NLP technique were provided.

As discussed above, to derive the categories for all attributes/aspects in the systematic map (Table 3), we used attribute generalization and iterative refinement and marked the categories as we were finding them in the papers. For any category that appeared in at least five papers, we created a new category in the corresponding set, otherwise, we added them to a respective category called "Other".

### 3.5 DATA EXTRACTION PROCESS FOR SYSTEMATIC MAPPING AND REVIEW

Once the systematic map (classification scheme) was ready, each of the researchers extracted and analyzed data from the subset of the papers (assigned to her/him). We included traceability links and added them to the extracted data to the exact phrases in the papers to ensure that we would suitably justify how we made each classification. For effective and efficient data extraction, we also used our recently-reported experience-based guidelines for this purpose [49].

Figure 6 shows a snapshot of our online spreadsheet that we used to enable collaborative work and classification of papers with traceability links (as comments). This snapshot shows the data for RQ 1.1 (Contribution type) in which one of the researchers has placed the exact phrase from the source as the traceability link to facilitate peer reviewing and quality assurance of data extractions.

**Figure 6- A screenshot from the online repository of papers (https://goo.gl/ZmuqZK).**

After all researchers finished data extractions, we conducted systematic peer reviewing in which researchers peer reviewed the results of each other's analyses and extractions. In the case of disagreements, we conducted discussions to reach a consensus. We conducted this process to ensure high quality of the extracted data and our results. Figure 7 shows a snapshot of our discussions during the peer reviewing process.

**Figure 7- A snapshot showing how peer reviewing of the extracted data was conducted**

## 4 RESULTS

This section presents results of the study's RQs. The section is structured according to the three groups of RQs:

- Group 1–Common aspects in all review studies (Classification of studies by contribution and research method types)
- Group 2-Technical issues specific to the topic (NLP-assisted software testing)
- Group 3-Specific to empirical and case studies

### 4.1 GROUP 1-COMMON TO ALL SLM STUDIES

We present the results for the RQs regarding group 1.

#### 4.1.1 RQ 1.1: Classification of studies by contribution types

Figure 8 shows the classification of studies by contribution types (facets). We can see that the majority of studies in this area has contributed methods or tools, besides only a few other contribution types (e.g., models or metrics). Note that as we discussed in the structure of the systematic map (Section 3.4 Table 3), since each study could have multiple contribution types, we could assign a study to more than one category in Figure 8.

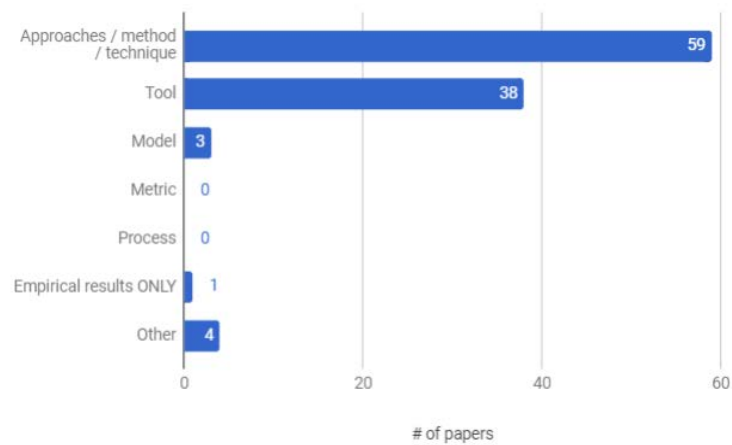

**Figure 8-(a): Classification of studies by contribution types**

We grouped approaches, methods, and techniques in one category, since their concepts are similar. They all implied a "way" to conduct an NLP-based analysis. 59 papers (~88% of the pool) contributed NLP-based approaches/methods/techniques to assist software testing. As we can see in Figure 8, this group is the largest category of the pool by contribution types. We will review various characteristics of NLP-based techniques in Section 4.2.

38 papers (~57% of the pool) presented tools to automate the presented approaches. Recall from Section 3.2 that we have a specific RQ (RQ 2.6) to review tool support in this area, and we will discuss it in Section 4.2.6.

Three (3) papers (4% of the pool), [P12, P13, P62], presented models to support NLP-assisted software testing. [P12] presented a specific model, named Use-Case Test Models (UCTMs). These models were generated from use-case specifications by NLP, and were then used to generate test cases. [P13] also focused on automatic generation of test cases from NL, and for this purpose, it also defined a specific model, named CSP (Communicating Sequential Processes) test models. The approach translated use-case documents into a formal representation in the CSP process algebra. Last, [P62] presented a language model which assigns a score to a string, reflecting its "likeliness" to occur in natural language. This score can then be utilized for search based structural test input generation.

The contribution of one paper [P4] was empirical results only. That paper reported an industrial study to analyze the performance of three existing NLP-based test-case prioritization techniques in the context of 30 industrial projects. No papers contributed metrics or processes in this context.

**4.1.2 RQ 1.2: Classification of studies by types of research methods**

Figure 9 shows the cumulative trend of mapping of studies by research facet. As we can see, a large amount of papers (40 papers, ~60%) were weak empirical studies, followed by 15 papers presenting solution proposals, and 12 papers with strong empirical studies. There were no "experience" papers.

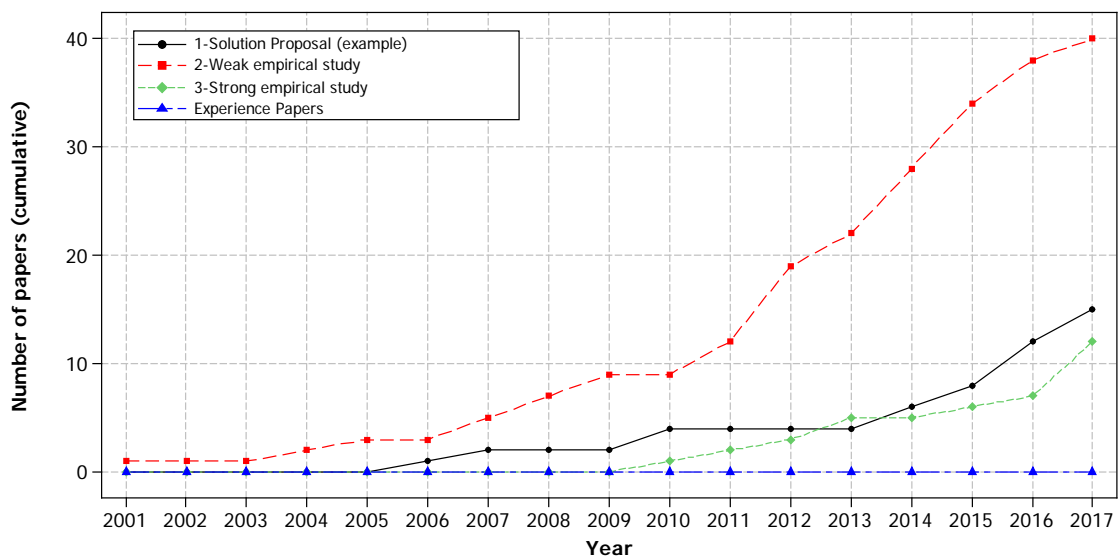

**Figure 9--Cumulative trend of mapping of studies by research-method types (n=67 papers)**

Since strong empirical studies are the most rigorous studies in this context, we have allocated a group of four RQs specific to them (RQ 3.1 …3.4) which we will review in Section 4.3.

**4.2 GROUP 2-SPECIFIC TO THE TOPIC (NLP-ASSISTED SOFTWARE TESTING)**

Group 2 of the RQs are more "technical" and specific to the topic of NLP-assisted software testing. We address the RQs under this group next.

**4.2.1 RQ 2.1-Type of NLP approaches used to assist software testing**

We classified the types of NLP approaches used to assist software testing. We used the classification of NLP approaches provided in the NLP community (Section 2.1) to assess this aspect. Figure 10 shows the breakdown. As we can see, the papers have used approaches from all three categories (morphologic, syntactic and semantic approaches) [2]. POS is the most widely used approach as perhaps it is the simplest (or most basic) approach. In general, semantic approaches are more sophisticated than morphologic and syntactic approaches [8].

Among the six "Other" semantic approaches, examples are: Cosine WordNet Similarity, used in [P17]; Language Extended Lexicon (LEL) analysis, used in [P19]; and case-grammar theory [50], used in [P28].

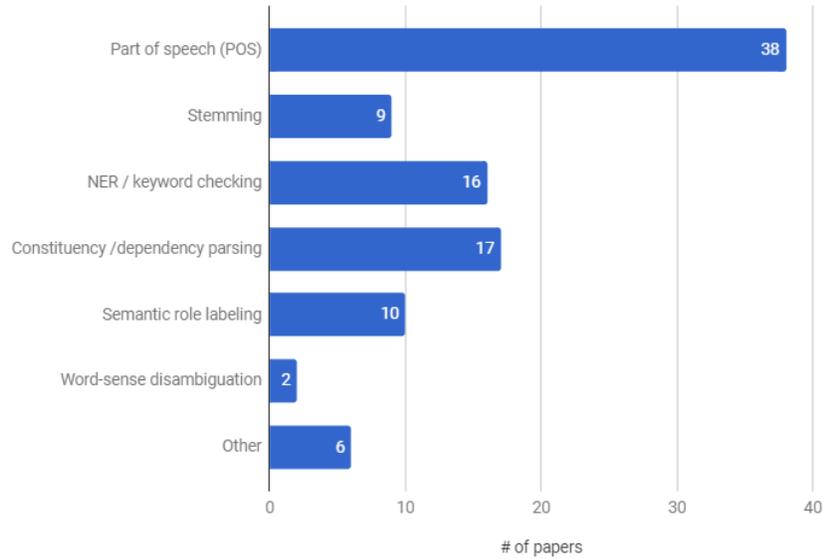

**Figure 10--Type of NLP approaches used to assist software testing**

### 4.2.2 RQ 2.2-Exposure level of the NLP approaches (algorithms) in the papers

We were keen to know the extent to which each paper presented details of the NLP aspects (algorithms). This RQ is important since we found that, while some papers exhibited (almost) all details of the presented NLP algorithm (in-depth), some papers discussed the presented NLP algorithms in a (very) shallow manner (almost no details). Our motivation for this RQ is that if a researcher or a practitioner wants to implement an NLP algorithm which is presented in a paper, s/he would need (almost) all algorithmic details (e.g., the NLP "pipeline" [2]) to develop it; otherwise, s/he cannot implement or use it.

As we showed in the systematic map (Table 3), we used a 3-point Likert scale for the exposure level: (1) very shallow exposure (almost no details), (2) average exposure (few details), and (3) in-depth exposure (most details). Figure 11 shows the breakdown. It is quite disappointing to see that a considerable amount of papers (32 of 67; 48%) had very shallow exposure to the NLP aspects (almost no details), and thus if a researcher or a practitioner wants to implement the NLP approaches in those papers, it will not be easily possible for her/him. 23 and 12 papers showed "average" and "in-depth" exposure levels respectively, by providing some or most of the details.

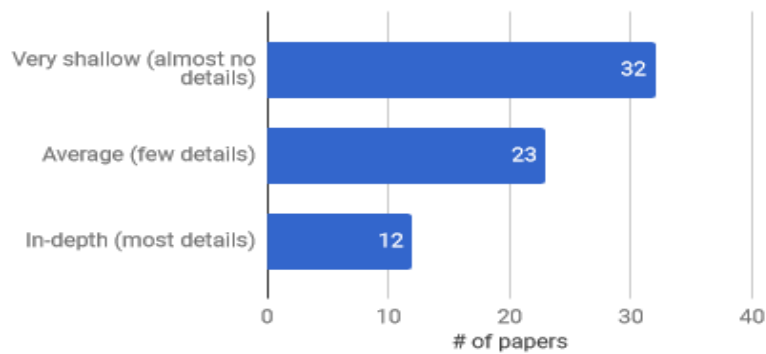

**Figure 11-- Exposure level of the NLP aspects (algorithms) in the papers**

### 4.2.3 RQ 2.3-Type of input NL requirements

Different papers considered different types for their NL input requirements. The different NL requirement formats provided an interesting insight into the practical prerequisites upon which NLP techniques could be applied. Based on the data presented in the papers, we categorized input formats as follows: (1) unrestricted (uncontrolled) NL, (2) restricted (controlled) NL; and (3) other. In total, 25 papers (37%) fell within the first category, whereas half (52%) of papers (35) required a restricted NL format. Seven (7) papers fell in neither of the above categories: [P7], for instance generated test cases from Java source code, and not NL requirements; [P62] created a customized controlled natural language for use case specifications.

An internal sub-categorization of (1) and (2) might prove useful to gain a better understanding of the difference between free NL and restricted NL requirement inputs. While most of the 25 papers generally used unrestricted NL requirements as inputs, some papers added small amendments to the freedom of NL. [P18], for instance, used unrestricted NL, however, requirements were required to follow a relatively well defined grammatical structure, called *action phrase* and *predicate phrase*. The restriction itself does not result in a controlled language, as the full flexibility of NL can still be used, but rather serves as a means to enable higher precision in deriving test cases. Similarly, [P42] used a predefined ontology specific to the context at hand (nuclear systems), to convert a textual document into an explicit system model for scenario-based test-case generation. Again, the ontology only imposed certain restrictions on the NL, but did not modify the underlying structure in its entirety.

Restricted NL formats also came in different varieties, reflecting the underlying approach and purpose of the succeeding NLP technique. Most papers focused on restricting writing of use cases according to a predefined structure, which should facilitate the NLP approach afterwards. [P9, P12], for instance, described a procedure named Restricted Use-Case Modeling (RUCM). Similar in concept, [P10] used the so-called Restricted User Story (RUST) to limit the NL input format. That study created the restricted NL inputs using template-like restrictions, which limited writing of use cases to a certain format. Tackling requirements representation from a logics point of view, [P6, P31] opted for an input presentation called *Courteous Logic*, where the information stored in the courteous logic predicates is used to automatically generate the test cases. Such a representation requires a high degree of abstraction when creating the use case descriptions, but can represent logical relationships more clearly.

**4.2.4 RQ 2.4-Intermediate model types**

In many papers, the presented approaches did not directly transform NL requirements into test cases. To facilitate transition by NLP of requirements to test-case generation, some papers often utilized an "intermediate" model type. Such a model functions as both the *result* of the NLP procedure and as the *input* model from which test cases were then derived. Our pool of papers exhibited several types of intermediate models, all with the purpose of enabling test-case generation.

Of the 33 papers which mentioned an intermediate model representation, six (6) papers used UML state machines, as they model the behavior of the software and represent both the control flow and dataflow [P2, P9, P23, P34, P39, P49]. State-machine-based testing is also popular in model-based testing.

Other intermediate model types mentioned were: communicating sequential processes (CSP), which is a type of behavioral model [P13, P22], activity diagrams [P14, P19], also Petri-Nets [P44, P45]. 15 other papers generated other specific model types, e.g., the so-called Restricted Test-case Models (RTCM) [P3,P35], object constraint language (OCL) [P8] and action target data (ATD) tuples [P58].

**4.2.5 RQ 2.5-Tool support (tools presented)**

Another important part were the tools which were developed and presented in the papers. In total, 36 of the 67 papers (54%) included in their description the development of a (research-prototype) tool, which could aid in automating the presented test-case generation approach. Moreover, we analyzed their online availability as per information provided in the papers. We found that, unfortunately, only three of those 36 tools (8%) were available for download: a tool named Toradocu [P11] (github.com/albertogoffi/toradocu), as well as a tool named C&L [P19] (pes.inf.puc-rio.br/cel), and the supporting prototype tool TORC [P55] (http://sourceforge.net/projects/torc-plugin).

Another source in the pool, [P14], which was an MSc thesis, also presented a tool (an Eclipse plug-in) for automatic generation of test cases using NLP. The thesis did not, however, provide any information about the online availability of the tool, but provided in its appendix the tool's installation manual.

In recent years, discussions have emerged in the research community about the importance of making research tools available, as this would imply various benefits, e.g., reproducible research [26-28, 51]. A 2010 paper [52] in the Communications of the ACM, expressed this issue as: "*Software code [behind research papers] can provide important insights into the results of research, but it's up to individual scientists whether their code is released-and many [scientists] opt not to*". This issue has been the subject of debate in the scientific community in large for many years, e.g., a scientist casted his opinion as: "*Freely provided working code – whatever its quality – improves programming and enables others to engage with your research*" [53]. A 2010 paper in Nature, entitled "*Publish your computer code: it is good enough*" [53], interviewed researchers about their reasons not to publish their research tools and here are some example replies: "*It is not common practice. People will pick holes and demand support and bug fixes. The code is valuable intellectual property that belongs to my institution. It is too much work to polish the code.*"

### 4.2.6 RQ 2.6-NLP tools used

27 of the 67 papers mentioned the name (s) of the NLP tool(s) used. We would imagine that all papers had to use NLP tools to automate the NLP analysis, but not all papers explicitly mentioned the tool names. Table 4 summarizes the list of NLP tools used in the papers.

**Table 4: NLP tools used in the studies**

| NLP tools | Number | references of the studies using the tools |
|---|---|---|
| 1. Stanford parser | 10 | [P7, P9, P11, P14, P17, P23, P39, P55, P56, P58, P63] |
| 2. NTLK | 3 | [P5, P38, P53] |
| 3. CNL parser | 2 | [P29, P30] |
| 4. CaboCha | 1 | [P1] |
| 5. MeCab | 1 | [P1] |
| 6. LTP | 1 | [P4] |
| 7. SCOWL | 1 | [P7] |
| 8. GATE | 1 | [P12] |
| 9. graph-based algorithm | 1 | [P23] |
| 10. LG parser | 2 | [P27, P58] |
| 11. Charniak parser | 1 | [P34] |
| 12. Shift maximum entropy parser | 1 | [P24] |
| 13. Unnamed tool | 2 | [P41, P64] |
| 14. Python NLP toolkit | 1 | [P51] |
| 15. Python library *genism* | 1 | [P52] |
| 16. Alchemy API | 1 | [P53] |
| 17. Tropes | 1 | [P66] |

A total of 17 different NLP tools have been used in the pool of papers, varying both greatly in their capacities and in their degree of recognition in the field. Most notably in this regard is the Stanford Parser [13] and the Natural Language Toolkit (NLTK) [30], which offer a broad variety of NLP-related functionalities. For the former, ten usages were reported, whereas the NLTK was used in three papers.

The 15 remaining tools were, at most, mentioned in two papers, or exhibited no repeated use. Partly, we can attribute this to contextual circumstances: [P4], for instance, used a Chinese platform called LTP to conduct their NLP processes, as the use cases were written in Chinese (see Section 4.2.7). This is also the case for [P1], which uses two parsing tools called CaboCha and MeCab, to parse Japanese NL requirements. In [P23], the NLP approach was conducted via an adaptation of the graph-based algorithm for word-sense ambiguity proposed by [54]. The actual adaptations, however, were not detailed. The same can be said about [P41] which only stated that they utilized an NLP parser taken from [55]; similarly, [P64] only states that is uses an unnamed parser. [P29, P30] (two subsequent papers by the same authors about the same tool) used a so called Controlled Language (CL) parser, the origin of which was not defined.

### 4.2.7 RQ 2.7- Support for different (natural) languages in requirements

As expected, the dominant language used for requirements specification was English, used by 64 (96%) of the 67 selected papers. Two papers [P15, P37] focused on automatic test-case generation from Japanese specification documents. In a similar manner, another paper [P4] described test-case generation from Chinese requirements. Interestingly enough, the papers themselves were written in English, which made the work accessible to all the community.

### 4.2.8 RQ 2.8-Output of the technique (types of generated test artifacts)

Unsurprisingly, for the majority of papers (52 papers), the resulting test artifacts were test cases. For example, [P7] explored the automated discovery of valid strings, and used them as test input data. [P16] automated the text input generation used for mobile testing. Many different forms of test-case generation were conducted, but they only differed conceptually, as in the end, all methods produce test cases in one form or another.

One paper [P11] used NLP to generate test oracles for exceptional behaviors, i.e., not test cases with concrete expected outputs, but mechanisms to decide whether test behavior is exceptional or not.

As a third distinction, we introduced a category called "artifacts in support of testing", which were the class of approaches that would generate artifacts in "support" of testing, but do not produce per se test cases themselves. 15 papers fell under this category. [P9], for instance presented an automated approach to generate state machine diagrams from use cases. This approach is especially interesting, as state machine diagrams can be utilized to generate test cases (see Section 4.2.4). [P33], on the other hand, proposed another approach to obtain models in support of testing from natural-language-like functional specifications. The focused domain was control software for passenger vehicles. The model format was in propositional logic and temporal relations. [P62] created readable string test inputs using a natural language model.

We also had an "Other" category for the outputs of techniques (types of generated test artifacts). One paper in the pool did not generate any test artifacts [P4], but instead was an industrial study of NLP-based test-case prioritization.

### 4.3 GROUP 3-SPECIFIC TO EMPIRICAL STUDIES AND ALSO THE CASE STUDY OF EACH PAPER

In this section, we address RQ 3.1 (Research questions studied in the empirical studies), RQ 3.2 (Scale (size metrics) of the case study) and RQ 3.3 (Methods used to evaluate the NLP approaches and empirical evidence) respectively.

#### 4.3.1 RQ 3.1-Research questions studied in the empirical studies

To assist us and readers (e.g., younger researchers) in exploring the type of research questions asked in this area and for pursuing interesting future research directions, we extracted the list of research questions (RQs) in the included studies. In total, 11 (of 67) studies raised and addressed 29 RQs. We extracted these RQs and have done a thematic coding on their goals, as listed in Appendix 2. Through the thematic coding from the RQs we identified the following five categories: (1) RQs assessing the feasibility of the presented NLP techniques, (2) RQs assessing the precision or efficiency of the techniques, (3) RQs investigating how much testing effort can be reduced by the presented NLP techniques, (4) RQs comparing the presented NLP technique to other techniques, and (5) other types of RQs.

An example for an RQ which assessed the efficiency of the investigated approaches is the following: "*What is the performance overhead incurred by our tool?*" [P16]. An example for an RQ which assessed the understandability of the generated test artifacts is the following: "*Are the test cases readable and comprehensible?*" [P31].

Our initial thematic coding of the combined set of 29 RQs studied in 11 (of overall 67) papers helps other researchers choose interesting and relevant future research directions in the area of NLP-assisted software testing. Furthermore, it also forms an important basis for a more in-depth systematic literature review (SLR) in the future to assess and compare the strengths and limitations of different NLP-assisted software testing approaches.

#### 4.3.2 RQ 3.2-Scale (size metrics) of the case study

We assessed the scale of the reported case studies by three size metrics: (1) the number of systems (or just examples), (2) the number of requirements artifacts processed by each NLP approach, and (3) the number of test cases produced by each NLP approach.

Figure 12 shows the histogram of the number of systems (or examples) used for evaluations in the papers. 44 sources evaluated their presented approaches on one SUT (or example). Only 22 papers worked on more than one SUTs. [P4] and [P16] conducted extra-ordinary evaluations by trying their approaches on 30 and 50 SUTs, respectively. In [P4], the authors used 15,059 test cases from 30 mobile applications to train the NLP-based machine-learning approach. In [P16], the NLP-based approach was evaluated with 50 iOS apps, including popular apps such as Firefox and Wikipedia. In [P52], the authors applied their approach on 100 real-world web (HTML) "forms".

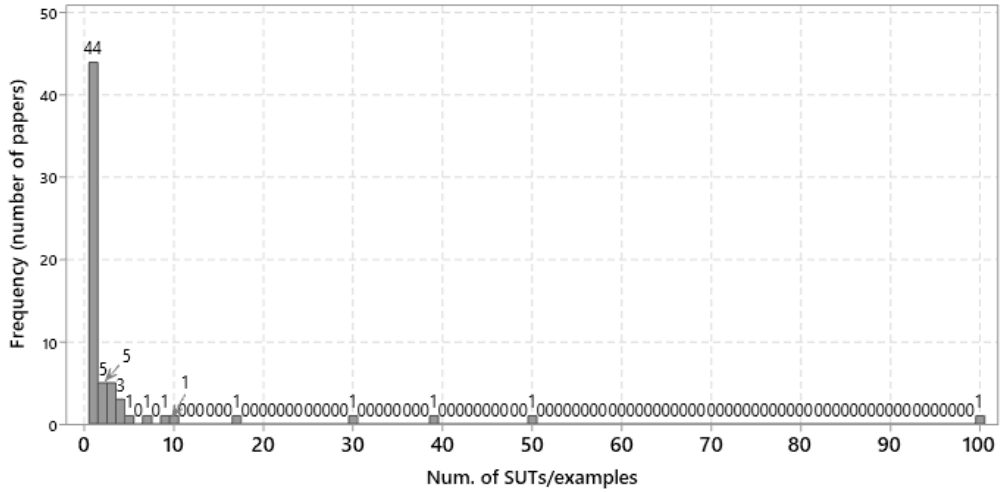

**Figure 12 — Histogram of the number of systems (or just examples) used for evaluations in the papers**

For the other two size metrics (the number of requirements artifacts processed and the number of test cases generated by each NLP approach), Figure 13 shows the X-Y scatter plot of those data. 11 studies reported these two metrics.

[P27] evaluated the presented approach to test-case generation by applying it to 1,841 requirements, which is the highest overall among all papers, from seven SUTs, and out of those requirements, the paper generated 1,582 test cases via NLP. We can see in Figure 13 that there is a wide range in the number of evaluations. Most of these 11 papers have evaluated the approaches on rather small-scale cases, i.e., less than 200 requirement items processed and less than 200 test cases generated.

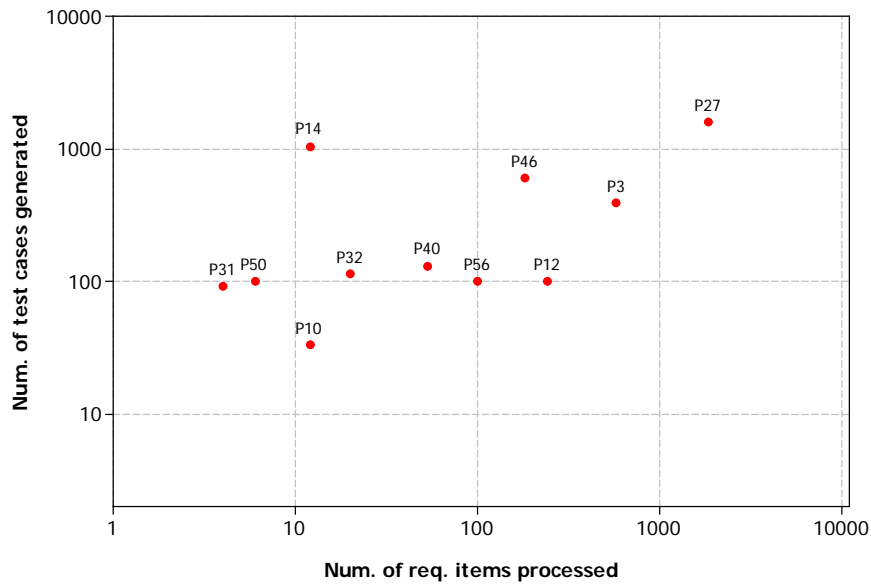

**Figure 13 — Scatter-plot of the number of requirements artifacts processed by and the number of test cases generated in the papers**

### 4.3.3 RQ 3.3-Methods used to evaluate the NLP approaches and empirical evidence

Papers have used various methods to evaluate the proposed NLP approaches. Based on our past experience in using grounded theory in systematic reviews, e.g., [40], we conducted a qualitative coding [48] of the evaluation methods used in the papers and iteratively developed the following categories of applied evaluation methods: (1) Proof of concept (feasibility), which is the most basic evaluation approach, (2) Accuracy in test generation (e.g., using metrics such as precision and recall), (3) Reduction in (test-case generation) effort, (4) Coverage measurement, (5) Mutation testing, (6) Generating additional test cases compared to manual testing, (7) detection of real faults, and (8) "Other". Figure 14 shows the classification results of the evaluation methods. It could be that a paper would use more than one evaluation method. We provide examples of each category next.

After evaluating the NLP approach in [P1], the researchers mentioned that: "*the test-case generator was able to extract the test cases in the accuracy at the same level as doing manually, and it greatly reduced the required time for processing*". Thus, we assigned it to two categories: Accuracy measurement and Reduction in effort. We also assigned [P5] to the same categories since it discussed its evaluation method as follows: "*To compare results to how a human tester performs versus the intelligent automated tester, a subset of errors [defects] that human testers found were organized into two groups: (1) errors [defects] that the automated tester should identify; and (2) errors [defects] that only human testers can find. After analyzing the subset of errors that the automated tester should identify, the automated tester found all of the errors human testers found plus four times more errors*". [P5] also mentioned that: "*The automated tester takes less than a day to run for a complete regression test, while a similar regression test for humans takes about two months with a team of seven engineers*".

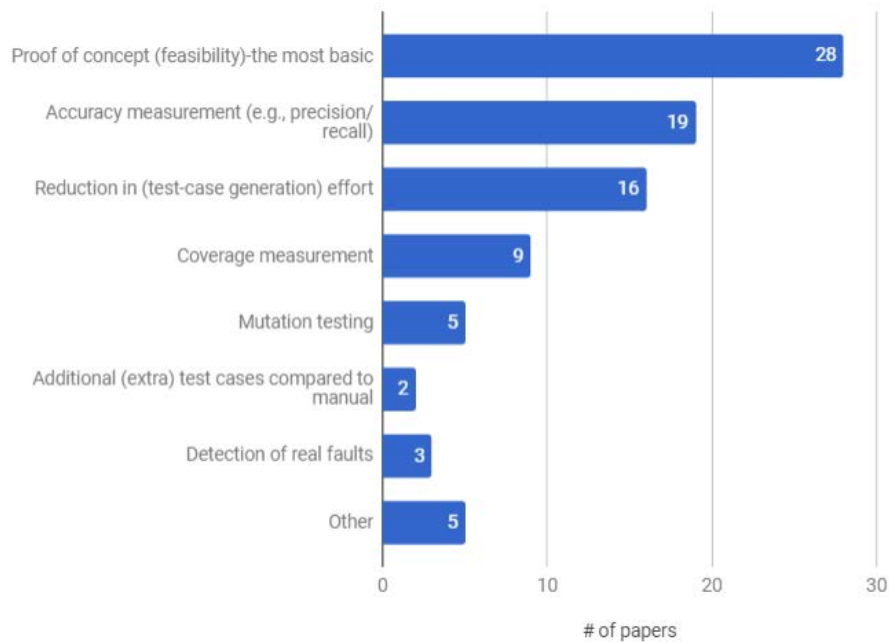

**Figure 14- Classification of methods used to evaluate the NLP approaches**

Proof of concept is the most basic (simplest) type of evaluation method in which only feasibility of the proposed NLP approach, often using a "running" example, was reported without a comprehensive case study, e.g., [P13] which mentioned: "*We illustrate our tool and techniques with a running example*".

Accuracy measurement was the second most common approach, which compares an NLP-generated test suite to a previously available test set (usually derived manually). Precision and recall are often calculated to evaluate the NLP-generated test set against the baseline test set. This evaluation approach was applied in 19 papers, which we review in RQ 3.4 (Section 4.3.4).

For instance, [P47] assessed how much effort the proposed approach saved by measuring improvement in test-case generation productivity, which increased from 77 requirement-lines/man-day in a manual approach to 110.50 requirement-lines/man-day using the NLP-based approach. We shall note that this evaluation approach should be complemented by an effectiveness measure, for instance precision and recall. To have a more precise NL-based test-case generation, one may need to use a more sophisticated technique and put in quite a lot of effort. Assessing reduction in (test-case generation) effort is another common evaluation method, which was applied in 16 papers.

Nine papers applied coverage-related evaluation methods. For example, [P31] assessed whether test cases cover the use-case scenarios exhaustively. [P50] compared the number of scenarios covered by manually-derived and automatically-generated test cases.

Detection of artificially-injected faults (mutation testing) was another evaluation method.

Two papers [P22, P23] found that their proposed NLP approach could result in additional (extra) test cases compared to the manually-generated approach. The evaluation methods in two other papers [P34, P40] included detection of real faults.

Several other papers [P11, P31, P36, P38, P62] used "Other" evaluation methods, all of which were "non-functional" in nature. [P31] evaluated readability and comprehensiveness of the system-generated test cases. [P36] evaluated the scalability of the automated test-generation approach. [P38] evaluated maintainability, reusability, and modularity of the system-generated test cases. Last, [P62] evaluated the capabilities of a language model for test input generation.

### 4.3.4 RQ 3.4-Accuracy (precision) of the approaches

For readers who could potentially consider applying an NLP-based test-case generation approach, accuracy (precision) of the approach is important since they would want to know how effective the approach is (e.g., what ratio of manual test cases could be generated by an automated approach?) A related approach is to measure the similarity between generated and manual test cases. Accuracy of the approaches was reported in 18 papers. We extracted the reported accuracy scores and show their histogram in Figure 15.

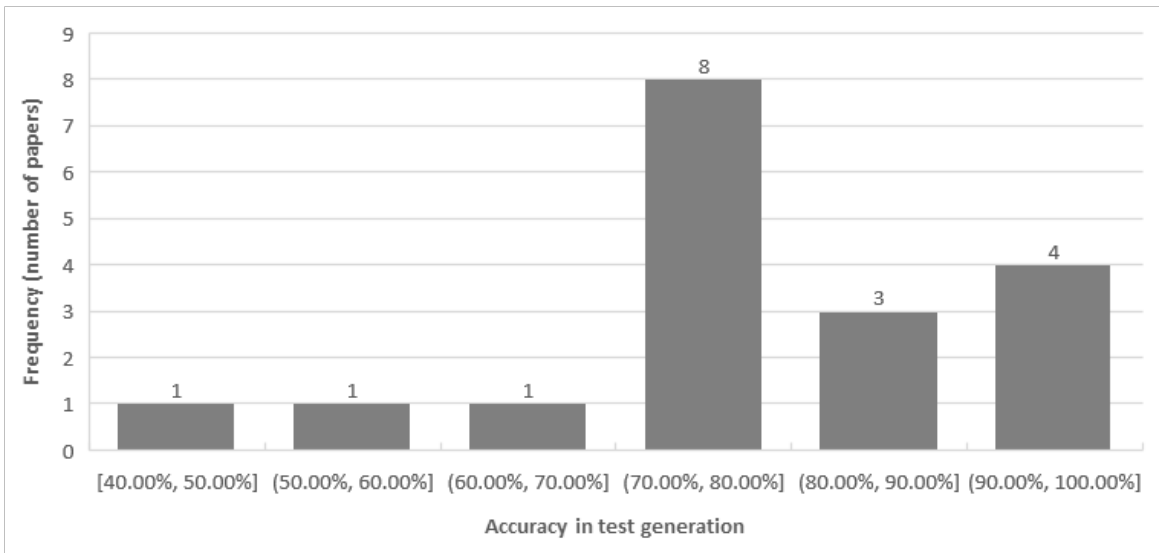

Figure 15- Accuracy of NLP-based test-case generation approaches in the pool under study

The lowest accuracy score was reported in [P7] in which automated generation of valid test strings using regular expressions and NLP was reported. [P7] reported that the ratio of average percentage of valid values generated using NLP was between 36-40% (we considered 40% for that paper in Figure 15). Accuracy of 100% was reported in two studies [P3, P43]. [P3] evaluated the presented approach with four case studies. Results showed that the approach was able to correctly process "all" (100%) the requirements in the case studies. [P43] compared the automatically-generated test set to manually-written test cases, and found that 85% of the manual test cases were generated with a precision of 100%.

As we can see in Figure 15, most of the reported accuracy scores lie in the range of 70%-90%. For example, in [P17] which was a work in the context of behavior-driven development (BDD), given a set of behavior descriptions, the approach was "*able to accurately convert about 73% of the 80 behavior descriptions into step definitions correctly*".

## 5 DISCUSSIONS

We provide a summary of findings and implications of our results. We then assess benefits of this review study, and discuss potential threats to validity.

### 5.1 BENEFITS OF THIS SYSTEMATIC MAPPING

Let us recall from Section 1 that this review study was conducted based on a real need that we had in our industrial projects. The authors and their collaborators have already started to benefit from the results of this review. In our ongoing collaborations with several industry partners in Turkey, Austria and the Netherlands in the area of software testing, our colleagues and we did not have an adequate overview of the literature and this review provided that. Thanks to our review study, we are currently assessing several existing NLP-based test techniques based on the review at hand for possible adoption/extension in our ongoing industry-academia collaborations.

To further assess the benefits of this review, we asked two test engineers from our industrial contacts (one in Austria and one in Turkey) to review this review paper and the online spreadsheet of papers, and let us know what they think about their potential benefits. Their general opinion was that a review paper like this article is an invaluable resource and can actually serve as an "index" to the body of knowledge in this area.

One of the practitioners provided the following feedback: "*The survey on NLP-assisted software testing contains many approaches which could be considered for an improvement of requirements-based testing in industry. The implementation of NLP approaches, dealing with generation of test cases from natural language requirements and semi-formal use-case models, would be beneficial in an industrial context. The article could foster its implementation in industry.*"

The other practitioner provided the following feedback: "*In our company, according to the effort/time logs that we have recorded for our past testing projects, our test engineers spend a large amount of effort on manually extracting test cases from requirements documents. It is interesting for me as a test manager to see that such a large number of techniques exists on NLP-assisted software testing. I would be interested to use this survey paper to review some of those techniques with the possibility of adopting/extending some of those techniques in our industrial context. Furthermore, I agree with the authors that many papers in general have low exposure of the NLP approaches (algorithms) in them. This severely decreases the chances of using those techniques in industry*".

Another important issue concerns the reported accuracy scores of the NLP-based test generation approaches. As we found in Section 4.3.4, the reported accuracy scores lie in the range of 70% to 90%, indicating that the approaches are quite effective in deriving relevant test cases, but we are still not at 100% accuracy level. For a wide industrial usage of NLP-based testing, practitioners would ideally require almost-100% accuracy, and thus, there is a need for more work. Also, as we saw in Section 4.1.2, there is limited industrial empirical assessments of the approaches in this area. More work is also required in that direction. We believe the above points are other interesting and useful contributions of this review paper.

Last but not the least, we should clarify that we have taken a first step (systematic classification of the literature) in this work, and a more thorough systematic literature review (SLR) is needed in future studies to assess and compare, objectively, the strengths and limitations of different NLP-assisted software testing approaches that we have classified in this initial SLM paper. Such a follow-up SLR could possibly be augmented with additional new empirical studies in which Systems Under Test (SUTs) are chosen, to which then the NLP-based approaches (from the papers) are applied. This could enable a critical assessment of the approaches under study. We believe this current work will provide inputs and guidance to such an extended study in future.

## 5.2 POTENTIAL THREATS TO VALIDITY

The main issues related to threats to validity of this literature review are inaccuracy of data extraction, an incomplete set of studies in our pool due to limitation of search terms, selection of academic search engines and researcher bias with regard to exclusion/inclusion criteria. In this section, these threats are discussed in the context of the four types of threats to validity based on a standard checklist for validity threats presented in [56]: internal validity, construct validity, conclusion validity and external validity. Next, we discuss those validity threats and the steps that we have taken to minimize or mitigate them.

Internal validity: The systematic approach that has been utilized for source selection is described in Section 4. In order to make sure that this review is repeatable, search engines, search terms and inclusion/exclusion criteria are carefully defined and reported. Problematic issues in the selection process are the limitation of search terms and search engines, and the bias in applying exclusion/inclusion criteria.

Limitation of search terms and search engines can lead to an incomplete set of primary papers. In order to find all relevant sources, a systematic search approach using carefully-defined keywords were conducted, followed by manual search in the list of references of the initial pool and in web pages of active researchers in our field of study. To maximize the chances of including all relevant studies, we used two search engines: Google Scholar and Scopus.

Furthermore, as discussed in Section 3.2.1, in addition to our defined search string for Google Scholar, which was "*Software AND (test OR testing) AND (natural language OR NLP OR natural language processing)*", we experimented with other similar terms, e.g., "processing of plain language" instead of the last operand after "AND" in the above search string: "*(natural language OR …)*". In addition, to ensure maximizing our chances to including as many relevant papers as possible, we experimentally designed, for the Scopus database, an additional search string as: *( TITLE-ABS-KEY ( ( test OR testing ) AND ( "natural language" OR nlp OR "natural language processing" ) ) AND SRCTITLE (software) )*. The *SRCTITLE(software)* term limited the search scope to only venues (journal or conference names) which include the term "software", and is an effective way to search for software engineering papers, as also used in past studies, e.g., [35]. Searching by the latter search string retrieved a set of additional 26 candidate papers. Therefore, we believe that we have been able to populate an adequate and inclusive candidate paper pool for this study; and the rate of missing publications should be negligible.

Applying inclusion/exclusion criteria can suffer from researchers' judgment and experience. Personal bias could be introduced during this process. In the case of disagreements, we conducted discussions to reach consensus. Also, to minimize human error/bias, we conducted extensive peer reviewing to ensure the quality of the extracted data.

Construct validity: Construct validity is concerned with the extent to which the object of study truly represents theory behind the study [56]. Threats related to this type of validity in this study were suitability of RQs and categorization scheme used for the data extraction. To mitigate these threats, the RQs were discussed among the authors and the categorization schemes were derived from established categorization schemes if such schemes were available.

Conclusion validity: Conclusion validity of a literature review study is concerned with whether proper conclusions are reached through rigorous and repeatable treatment. All primary studies are reviewed by at least two authors to mitigate bias in data extraction. Each disagreement between authors was resolved with consensus among researchers. Following the systematic approach and described procedure ensured replicability of this study and assured that results of a similar study will not have major deviations from our classification decisions.

External validity: External validity is concerned with the extent to which the results of our literature review can be generalized. As we saw in Section 5.1, the collected papers contained a significant proportion of academic and industrial work which forms an adequate basis for concluding results useful for both academia and industry. Also, note that our findings in this study are mainly within the field of test-case generation from NLP requirements. We have no intention to generalize our results beyond this subject.

# 6 CONCLUSIONS AND FUTURE WORK

By classifying the state-of-the-art and the –practice, this survey paper mapped and reviewed the body of knowledge on NLP-assisted software testing. We systematically reviewed 67 papers in this area and classified them. By summarizing what we know in this area, this paper provides an "index" to the vast body of knowledge in this area. Practitioners and researchers who are interested in reading each of the classified studies in depth, can conveniently use the online Google spreadsheet at goo.gl/VE6FeK to navigate to each of the papers.

While we conducted in this work a basic level of synthesis for RQ 3.4 (accuracy of the approaches), a more through systematic literature review (SLR) is needed to assess and compare, objectively, the strengths and limitations of different NLP-assisted software testing approaches that we have classified in this initial SLM paper. Such a SLR could possibly be augmented with additional new empirical studies in which a few Systems Under Test (SUTs) are chosen, to which then the NLP-based approaches (from the papers) are applied. This would enable a critical assessment of the strengths and weaknesses of the approaches under observation. As we have taken the first step (comprehensive systematic classification of the literature) in this work, we leave the above in-depth systematic literature of evidence in this area to future works. We believe this current work will provide inputs and guidance to such an extended study.

It is the hope of the authors that practitioners would utilize various ideas discussed in this review and each of the 67 papers, and then report back to the community how each idea helped them to use NLP-assisted software testing in their projects. We also encourage practitioners to report their concrete challenges in the area of NLP-assisted software testing so that researchers can work on and solve those challenges.

## APPENDIX 1-LIST OF THE PRIMARY STUDIES REVIEWED IN THIS SURVEY

| | |
|---|---|
| [P1] | A. Takashima, H. Maruyama, Y. Lu, and T. Nakamura, "A Method for Extracting Test Cases from a Basic Design Document," *Joint Conference on Knowledge-Based Software Engineering,* pp. 243-251, 2010. |
| [P2] | R. Chatterjee and K. Johari, "A prolific approach for automated generation of test cases from informal requirements," *ACM SIGSOFT Software Engineering Notes,* vol. 35, pp. 1-11, 2010. |
| [P3] | M. Zhang, T. Yue, S. Ali, H. Zhang, and J. Wu, "A systematic approach to automatically derive test cases from use cases specified in restricted natural languages," *International Conference on System Analysis and Modeling,* pp. 142-157, 2014. |
| [P4] | Y. Yang, X. Huang, X. Hao, Z. Liu, and Z. Chen, "An Industrial Study of Natural Language Processing Based Test Case Prioritization," *IEEE International Conference on Software Testing, Verification and Validation,* pp. 548-549, 2017. |
| [P5] | C. Bischke, Y.F. Choy, E. Urhiafe, and J. Dibenedetto, "An Intelligent Tester: Automated Test Generation and Test Execution through Machine Learning and Natural Language Processing," *International Conference on Software Engineering and Data Engineering,* 2016. |
| [P6] | R. Sharma and K. Biswas, "Automated generation of test cases from logical specification of software requirements," *International* |

| | | |
|---|---|---|
| [P61] | B. Hois, S. Sobernig, and M. Strembeck. "Natural-language scenario descriptions for testing core language models of domain-specific languages." *International Conference on Model-Driven Engineering and Software Development*, pp.356-367, 2014. | |
| [P62] | S. Afshan, P. McMinn, and M. Stevenson. "Evolving readable string test inputs using a natural language model to reduce human oracle cost." *IEEE International Conference on Software Testing, Verification and Validation*, pp. 352-351, 2013. | |
| [P63] | A. Genaid. "Connecting user stories and code for test development." IEEE *International Workshop on Recommendation Systems for Software Engineering*, pp. 33-37, 2012. | |
| [P64] | F. Barros et al. "The ucsCNL: A Controlled Natural Language for Use Case Specifications", *International Conference on Software Engineering and Knowledge Engineering (SEKE)*, pp.250-253, 2011. | |
| [P65] | M. Schnelte. "Generating test cases for timed systems from controlled natural language specifications." *IEEE International Conference on Secure Software Integration and Reliability Improvement*, pp. 348-353, 2009. | |
| [P66] | Y. Madhavan, J.W. Cangussu, and R. Dantu. "Penetration Testing for Spam Filters." *Annual IEEE International Computer Software and Applications Conference*, vol. 2, pp.410-415, 2009. | |
| [P67] | C. Bertolini, and A. Mota. "Using Refinement Checking as System Testing." *Conferencia Iberoamericana de Software Engineering*, pp. 17-30, 2008. | |

## APPENDIX 2- RESEARCH QUESTIONS (RQs) RAISED AND STUDIED IN THE EMPIRICAL STUDIES

| Paper # | List of Research Questions (RQs) | Thematic coding of RQs | | | | |
|---|---|---|---|---|---|---|
| | | Feasibility | Precision | Reducing effort | Comparison to other techniques | Other |
| [P7] | • RQ 1-Does the use of regular expressions and web queries formulated by the knowledge extracted from the program identifiers result in producing valid string values? If yes, what is the precision?<br>• RQ 2-Which web search strategies are significant in finding valid string values?<br>   ○ RQ 2.1-Which combination of NLP techniques is more significant for processing identifier names?<br>   ○ RQ 2.2-Which method for obtaining regular expression is more significant in finding valid values?<br>• RQ 3-How effective is the approach compared to the other test data generation techniques for strings? | | RQ1 | | Comparison to other test data generation techniques (RQ3) | Combination of NLP techniques (RQ 2.1) |
| [P11] | • RQ 1-Do developers test exceptional behavior less than normal behavior?<br>• RQ 2-To what extent does Toradocu reduce the number of false positives that test input generation tools produce?<br>• RQ 3-Does Toradocu reveal faults in the implementation of exceptional behavior? | | | | | Fault detection of generated test cases |
| [P14] | • RQ 1-Can we generate test cases from user stories in an Agile software development work flow using the test-case generation tool?<br>• RQ 2-Does this tool save the tester's time and effort while improving the quality and coverage of the test cases? | Feasibility (RQ1) | | RQ2 | | |
| [P16] | • RQ 1-How effective is our automated approach compared to other automated input generation approaches for mobile testing?<br>• RQ 2-Does Word2Vec allow better results compared to using the RNN model only?<br>• RQ 3-What is the performance overhead incurred by our tool? | | RQ2 (better results) | | Comparison to other test data generation techniques | Performance overhead |
| [P17] | • RQ 1-Can we reduce the burden on programmers by utilizing information in the natural language, so that they don't have to hand-write the glue code? | | | Reducing time/effort of manual testing | | |
| [P24] | • RQ 1-From use case descriptions, can we generate concrete test procedures and test cases?<br>• RQ 2-Can we also take into account the variability and dynamicity/context changes of the DSPL in this generation? | Feasibility (RQ 1) | | | | |
| [P31] | • RQ 1-Are the test cases readable and comprehensible?<br>• RQ 2-Do you think these test cases cover the scenario exhaustively? | | RQ2 | | | Readability and comprehensibility of generated test cases |
| [P35] | • RQ 1-Do users find RUCM too restrictive or impractical in certain situations?<br>• RQ 2-Do the rules and template have a positive, significant impact on the quality of the constructed UML analysis models? | RQ1 | | | | Too restrictive or impractical (RQ1); quality of generated models |
| [P47] | • RQ 1-Does a platform like Text2Test (one that enables edit time monitoring) ensure stronger compliance of use cases to a set of pre-defined guidelines?<br>• RQ 2-Does a platform like Text2Test improve productivity of use-case authors?<br>• RQ 3-Does guideline-compliance of use cases ensure higher quality test cases in model based test generation? | | | RQ 2 | | Higher quality test cases in model based test generation |
| [P52] | • RQ 1-What is the effectiveness of the proposed approach in input topic identification and GUI state equivalence computation, comparing with conventional methods used in crawling-based web application testing?<br>• RQ 2-How much training data is required for the proposed approach?<br>• RQ 3-Can the proposed approach be used to improve the rule-based one? | | | | RQ1 | |
| [P62] | • RQ 1-Does the use of the language model as part of the fitness function improve the language model scores of the strings generated?<br>• RQ 2-Accuracy of judgements: Is there an improvement in the accuracy of evaluation of strings generated using the language model? That is, do the participants enter the correct expected outputs for a string input produced using the language model more frequently than for strings generated without the use of the language model?<br>• RQ 3-Time to make judgements: Is there a decrease in time for evaluating strings produced using the language model? That is, do human participants enter the correct expected outputs for strings generated using the language model more quickly than those generated without the use of the language model? | | RQ 2. Accuracy of judgements | Reduce human oracle cost (in paper title) | | |